\documentclass[useAMS,usenatbib]{mn2e}
\usepackage{amsmath}
\usepackage{amssymb}
\usepackage{graphicx}

\title{Kinematics of solid particles in a turbulent protoplanetary disc}

\author[]{Augusto Carballido$^{1}$,\thanks{E-mail: augusto@jpl.nasa.gov}
James M. Stone$^{2}$,\thanks{E-mail: jstone@astro.princeton.edu}
Neal J. Turner$^{1}$ \thanks{E-mail: neal.turner@jpl.nasa.gov}\\
$\rm ^1$ Jet Propulsion Laboratory, MS 169-506, California Institute of
Technology, Pasadena CA 91109, USA\\
$\rm ^2$ Department of Astrophysical Sciences, Princeton University, Princeton
NJ 08544, USA \\}

\begin{document}
\maketitle
\begin{abstract}
We perform numerical simulations of solid particle motion in a shearing
box model of a protoplanetary disc. The accretion flow is turbulent
due to the action of the magnetorotational instability. Aerodynamic
drag on the particles is modelled using the Epstein law with the gas
velocity interpolated to the particle position. The effect of
the magnetohydrodynamic turbulence on particle velocity dispersions is
quantified for solids of different stopping times $t_{\rm{s}}$, or equivalently,
different sizes. The anisotropy of the turbulence is reflected upon
the dispersions of the particle velocity components, with the radial
component larger than both the azimuthal and vertical components for
particles larger than $\sim $ 10 cm (assuming minimum-mass solar nebula
conditions at 5 AU). The dispersion of the particle velocity
magnitude, as well as that of the radial and azimuthal components, as functions of
stopping time, agree
with previous analytical results for isotropic turbulence. The
relative speed between pairs of particles with the same value of
$t_{\rm{s}}$ decays faster with decreasing separation than in the case
of solids with different stopping time. Correlations in the particle
number density introduce a non-uniform spatial distribution of solids
in the 10 to 100 cm size range. Any clump of particles is disrupted by the
turbulence in less than one tenth of an orbital period, and the
maximally concentrated clumps are stable against self-gravitational collapse.
\end{abstract}


\section{Introduction}
In order for planetesimals to form in a circumstellar disc, the
parent dust grains must meet a variety of kinematic conditions if they
are to grow in mass and size. The formation of kilometre-sized solid
bodies by gravitational instability requires, through the Toomre
criterion, that the dust spatial density be $\rho_{d}>\Omega^{2}/\pi
G$, where $\Omega$ is the disc angular velocity and $G$ the
gravitational constant. Such densities could be achieved in a
mid-plane dust layer after sedimentation has occurred. On the other
hand, growth by collisional sticking may be accomplished if the grain
relative speeds acquire certain values: experiments have shown that,
for micron-sized grains, collision speeds less than $\sim$0.2 m/s lead to aggregates with a
fractal geometry (Blum \& Wurm 2000), and above $\sim$ 1 m/s
disruption of the colliding agglomerates takes place.  

While binary collisions of the smallest dust grains are caused by
Brownian motion, other sources of relative velocities in a
protoplanetary nebula may come into play as the interaction between
solids and gas decreases. Differential sedimentation, radial drift and
turbulence can provide the necessary kinetic energies to drive the
collisional motion of solids, in addition to changing their spatial
distribution throughout the nebular disc. In this paper we focus on
the effect of a turbulent gas flow on solid particle velocities and spatial
arrangement. 

The formalism of V\"{o}lk et al. (1980) makes it possible to calculate the
velocity dispersions of particles in isotropic turbulence. The effect
of the fluctuating gas velocity on individual solids is approximated
by a sum over two types of turbulent eddies: those with spatial
frequencies $k$ and turnover times $t_{k}>t_{\rm{s}}$, where
$t_{\rm{s}}$ is the particle stopping time, and which merely advect
the particle; and those with $t_{k}<t_{\rm{s}}$, which decay
before the particle has had time to cross them. The result is a
velocity dispersion that decreases with particle stopping time as
$(\Omega t_{\rm{s}})^{-1/2}$, with a sharp steepening for particles
with $\Omega t_{\rm{s}}\sim$ a few times $10^{-1}$. Under minimum mass
solar nebula conditions at 5 AU, this corresponds to particle radii of tens of
centimetres. The method is nicely summarised and applied by Cuzzi \& Hogan (2003) 
to chondrule-sized objects. 

The accumulation of a large number of solids in small regions of the turbulent
protoplanetary disc can lead to feedback effects on the gas flow if
the solid mass density is greater than the local gas density. Such
particle agglomerations can also lead to dust growth if the associated
collision rates are conducive to sticking. One recently suggested clumping
mechanism is a streaming instability that arises due to the relative
motion between solids and gas in a Keplerian disk (Youdin \& Goodman
2005). The feedback
of solids onto the gas can generate exponential growth of particle
density perturbations as well as turbulence, and particle clumps
can develop overdensities up to a factor of $\sim$ 50 relative to the
mean gas density (Youdin \& Johansen 2007, Johansen \& Youdin 2007).  

The process of turbulent
concentration (Maxey 1987) has been explored by Cuzzi et al. (2001),
who use the ratio of local
particle density to its global average as the measure of a
multifractal distribution of concentrated chondrules. This
distribution resembles that of the dissipation of the turbulent kinetic
energy on the Kolmogorov scale (Chhabra 1989). The connection between
turbulent concentration and turbulent dissipation suggests that the
local structure of the chondrule concentration field may be independent
of flow properties. For larger particles, an accurate characterisation of their
number statistics in the turbulent circumstellar environment is still lacking.

The recognition that turbulence due to the magnetorotational
instability (MRI; Balbus \& Hawley 1991; Hawley, Gammie \& Balbus
1995, henceforth HGB) can provide the viscous
stresses necessary to allow inward accretion in discs, has prompted
investigations of different dynamical aspects of dust evolution in the
presence of magnetohydrodynamic (MHD) turbulence. Planetesimal
formation by gravitational instability was addressed by Johansen et
al. (2006, henceforth JKH), whose numerical studies concluded that groups of 
at least 1000 particles in their simulations could become
gravitationally bound, assuming specific values of disc column
densities (150 and 900 g cm$^{-2}$) and a ratio of disc scale height
to radius of 0.04. Fromang \& Nelson (2005) examine the accumulation
of individual solids due to vortical structures in a global
protoplanetary disc model. They follow the radial migration of up to
3000 bodies, whose drag interaction with the gas corresponds to
objects of approximately 1 m in size. Those particles able to migrate
inwards do so at different rates, with reductions in semi-major axes
ranging from a factor of 1.6 in approximately 50 orbits to a factor of
2.2 in 200 orbits. The particles that become trapped in vortices
maintain an approximate constant radial position until the end of the
simulation. 

In the following we present a study of the kinematics of
solid particles in a local MHD model of a protoplanetary disc. In Section 2
we describe the numerical method employed to model the accretion flow
and the particles. Section 3 contains results obtained from
measurements of particle velocities and spatial distribution. In
Section 4 we discuss their significance in the context of particle growth
in a protoplanetary nebula, and concluding remarks are presented in
Section 5.

\section{METHOD}
\subsection{Numerical Method}
We use a three-dimensional version of the ZEUS code (Stone \&
Norman 1992a;b) in which the shearing box model of HGB is
implemented. To simulate the solid particles, we have added a module
that solves their equation of motion, 

\begin{equation}
\bf{f}=-\frac{1}{\rm{t_{s}}}(\bf{v_{\rm{p}}}-\bf{v_{\rm{g}}})-\rm{2}\bf{\Omega}
\times \bf{v_{\rm{g}}}+ \rm{3}\Omega^{2} x\bf{\hat{x}}
\label{eq:partmotion}
\end{equation} 

\noindent where $\bf{f}$ is the force
per unit mass on the particle, the first term on the right-hand side
is the Epstein drag law for spherical particles, $t_{\rm{s}}$ is the
particle stopping time,
$\bf{v_{\rm{p}}}$ is the particle velocity, $\bf{v_{\rm{g}}}$ is the
velocity of the background flow at the particle's position, and $\bf{\Omega}=\rm{(0,0,\Omega)}$
is the angular velocity of the box. The last two terms of
Eq.~(\ref{eq:partmotion}) are the Coriolis and tidal forces,
respectively, that arise as a result of the non-inertial character of
the shearing box ($x$ is the distance in the radial direction $\bf{\hat{x}}$). The particle module takes as input the gas velocities $\bf{v}$
calculated by ZEUS on the faces of the computational grid cells. These
velocities are obtained by solving the ideal MHD
equations for the shearing box, together with an isothermal equation of
state,

\begin{equation} 
\frac{\partial \rho_{\rm{g}}}{\partial t} + \nabla \cdot (\rho_{\rm{g}} \textbf{v})=0
\label{eq:mhd1}
\end{equation}

\begin{equation}
\frac{\partial \textbf{v}}{\partial t} + \textbf{v} \cdot \nabla \textbf{v}=-\frac{1}{\rho_{\rm{g}}} \nabla \left( P + \frac{B^2}{8\pi}\right) + \frac{(\textbf{B} \cdot \nabla) \textbf{B}}{4\pi \rho_{\rm{g}}} -\rm{2}\bf{\Omega} \times \textbf{v} + \rm{3}\Omega^2x\bf{\hat{x}}
 \label{eq:mhd2}
\end{equation}

\begin{equation}
\frac{\partial \textbf{B}}{\partial t}=\nabla \times (\textbf{v}\times \textbf{B})
 \label{eq:mhd3}
\end{equation}

\begin{equation} 
P=\rho_{\rm{g}}c_{\rm{s}}^2
\label{eq:mhd5}
\end{equation}

\noindent where $\rho_{\rm{g}}$ is the gas density, $\bf{B}$ is the
magnetic field, $P$ is the gas pressure, and $c_{\rm{s}}$ is the
isothermal sound speed. The module
interpolates each velocity component to the position of
each particle inside the respective computational grid cell, using a
trilinear interpolation algorithm (Press et
al. 1992). Equation~(\ref{eq:partmotion})
 is solved via a second-order Runge-Kutta
method. The back reaction of the particles on the gas, inter-particle
interactions, and particle coupling to the magnetic field are not included. 

Despite the likely presence of dead zones in protoplanetary discs as a
result of non-ideal MHD processes (Gammie 1996), in this first study we
neglect the effect of resistivity and assume that our local disc model
represents an active region in the surface layers or at a large
orbital radius (Sano et al. 2000).    

We adopt the values used in HGB for the initial density, initial pressure, angular
velocity, plasma parameter and box dimensions: $\rho_{0}=1$, $P_{0}=10^{-6}$,
$c_{\rm{s}}=\Omega=10^{-3}$, $\beta=400$ and
$H\times 2\pi H \times H$ (where $H$ is the disc scale height), respectively. The initial magnetic field
is vertical but, unlike HGB, it varies sinusoidally in the radial direction, in such a
way that the net magnetic flux is zero. The grid resolution used is
$84\times 180 \times 84$ zones. 

Simple tests of the particle integrator involved the evolution of a single
solid particle in a zero-velocity background flow, with various values of
the stopping time $t_{\rm{s}}$ which, in the Epstein regime, is given
by (Weidenschilling 1977). 

\begin{equation}
t_{\rm{s}}=\frac{a\rho_{\rm{p}}}{c_{\rm{s}}\rho_{\rm{g}}}
\label{eq:stoptime}
\end{equation}

\noindent where $a$ is the particle radius and $\rho_{\rm{p}}$ the
particle solid density. Given an initial position
and velocity, the response time of the particle to the drag force was
found to adjust accurately to the expected analytical behaviour for a
range of values of the dimensionless stopping time $\Omega
t_{\rm{s}}$ between 2$\times 10^{-3}$ and 100.

\subsection{Simulations}
We performed two sets of simulations in which the shearing box is first evolved for 10 orbits (1 orbit=$2
\pi/\Omega$) to ensure that turbulence due to the MRI has
developed. At this time, $N$ particles are introduced in the
flow with random initial positions and with initial velocities set to
zero. In simulation set A, $N=80,000$, and the total number of particles
is divided in eight groups of 10,000 members. Each group is given a
specific value of $\Omega t_{\rm{s}}$. The values used were
100, 30, 10, 5, 2, 0.2, 0.02 and 0.002. The system was evolved for a further 100 orbits. 

Simulation set B, which was used to study the particle spatial
distribution, consists of four groups of $N=4,704,000$ particles,
with each group having one of $\Omega t_{\rm{s}}$=2, 0.2,
0.125 and 0.05. The code was run for a further 30 orbits.

\section{RESULTS}
\subsection{Viscous evolution of disc}
\begin{figure*}\label{fig:alpha}
\begin{center}
\includegraphics[width=0.85\textwidth]{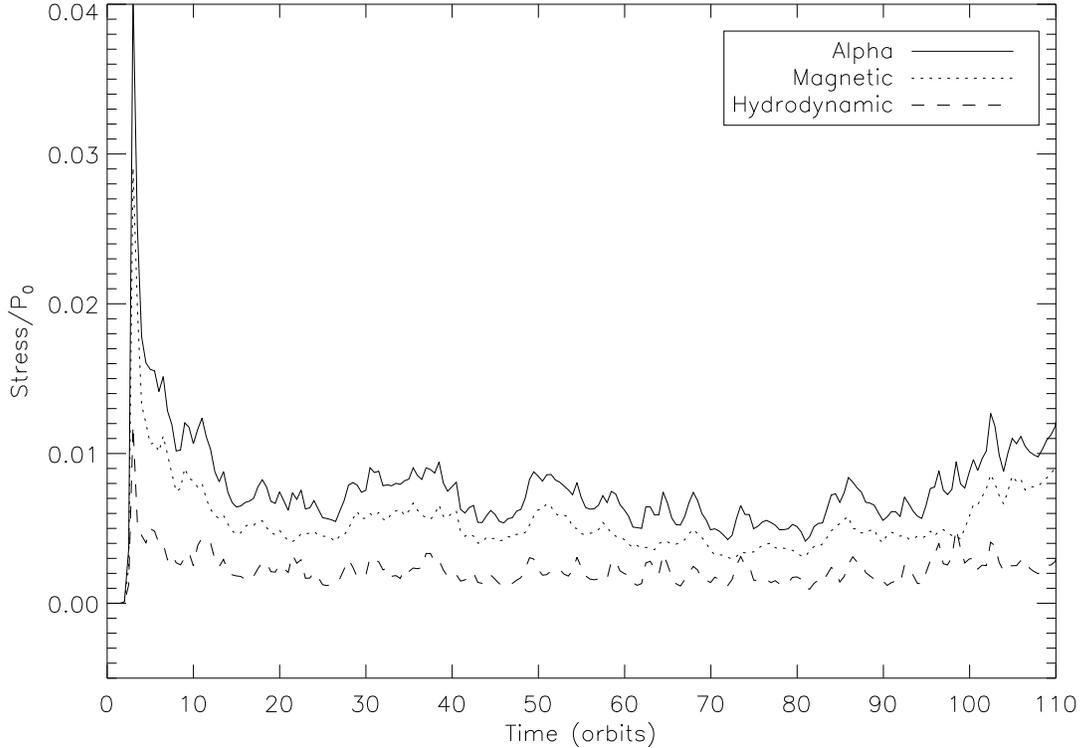}
\end{center}
\caption{Time evolution of the stresses in the local disc model. The stresses are normalized
to the initial gas pressure. The sum of the magnetic and hydrodynamic
stresses (solid line) provides an estimate of the viscous $\alpha$
parameter, whose average value is $8\times 10^{-3}$.  }
\end{figure*}

Figure 1 shows the evolution of the magnetic (\textit{dotted line}) and
hydrodynamic (\textit{dashed line}) volume-averaged stresses associated with our disc
model, during the longest run performed. The stresses are normalized
to the initial pressure. Their sum (\textit{solid
line}) gives an estimate of the volume-averaged $\alpha$ parameter of
classical disc theory. The average value of $\alpha$ is $8\times
10^{-3}$, which falls in the range $10^{-4}-10^{-2}$ inferred for T
Tauri stars (e.g. Hartmann et al. 1998).

\subsection{Velocity dispersions}
\begin{figure*}\label{fig:veldisp}
\begin{center}
\includegraphics[width=0.85\textwidth]{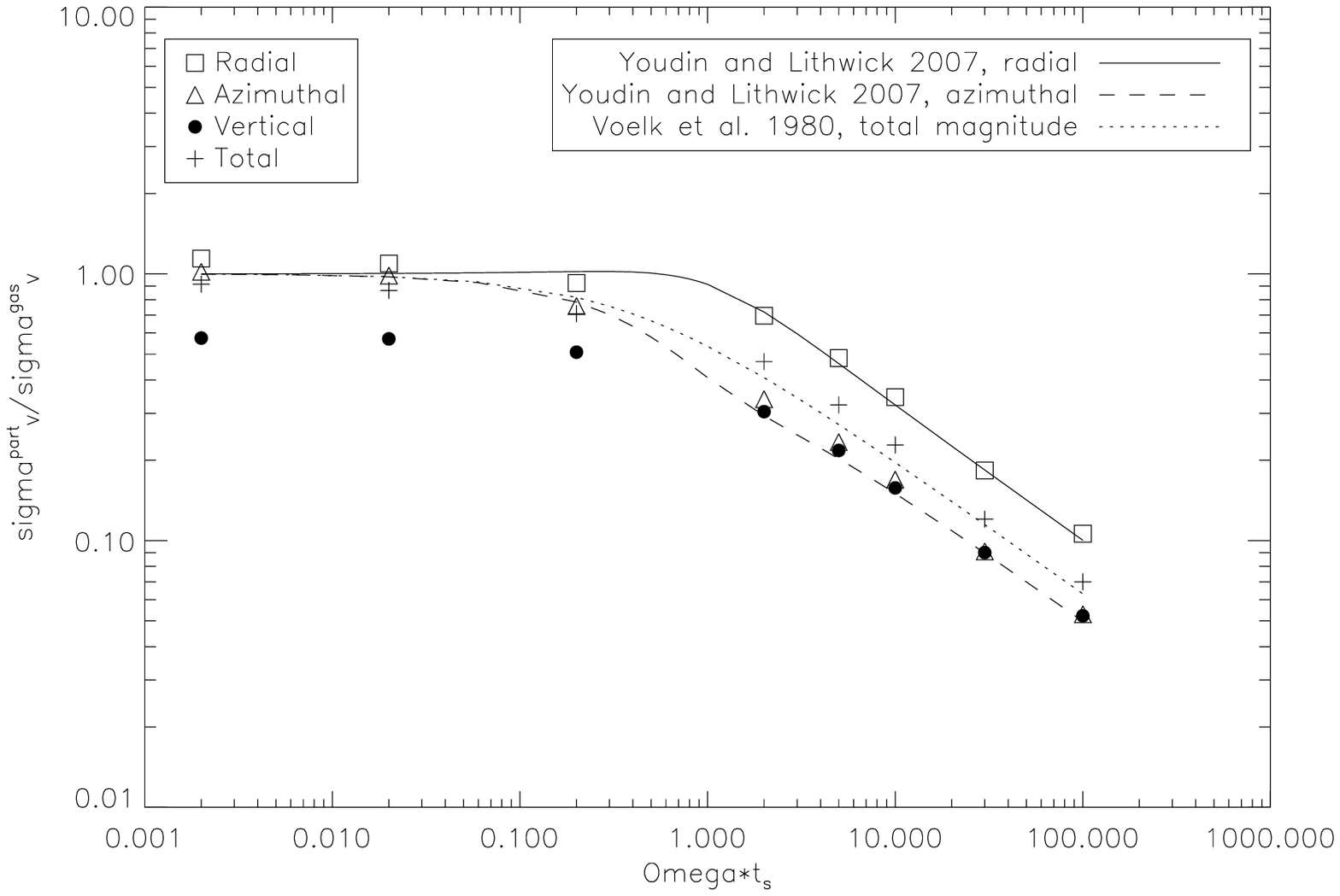}
\end{center}
\caption{Ratio of particle to gas velocity dispersion, as a function
of particle stopping time. Data shown correspond to the radial
(\textit{squares}), azimuthal (\textit{triangles}) and vertical
(\textit{circles}) components of velocity, as well as to the velocity
magnitude (\textit{plus signs}). The curves represent analytical
models derived for isotropic turbulence.}
\end{figure*}

Simulation set A has been used to calculate the time-averaged distributions of
the particle velocity
components. The resulting dispersions, $\sigma_{v_{i}}^{\rm{p}}$
($i=x,y,z$), normalised by the corresponding
gas velocity dispersions $\sigma_{v_{i}}^{\rm{g}}$, have been measured
as functions of
particle stopping time. To calculate this ratio, $S_{i}\equiv
\sigma_{v_{i}}^{\rm{p}}/\sigma_{v_{i}}^{\rm{g}}$, the gas velocities
were computed at the position
of the particles, at each snapshot and for each
value of $\Omega t_{\rm{s}}$, and the
effect of shear was subtracted from the particle and gas
$y$-velocities. The ratios $S_{i}$ thus obtained were then averaged over the
last 10 orbits of the simulation, a period in which the largest
particles (those with $\Omega t_{\rm{s}}$=100) have reached
dynamical equilibrium with the gas.

The results are summarised in Fig. 2, where the solid-to-gas ratios
of the $x$ (\textit{squares}), $y$
(\textit{triangles}) and $z$ (\textit{filled circles}) components of the
velocity dispersions are plotted. Also shown as plus signs are the ratios
$S_{v}\equiv \sigma_{v}^{\rm{p}}/\sigma_{v}^{\rm{g}}$ of the
velocity magnitude dispersions
$\sigma_{v}^{\rm{p,g}}=\sqrt{\langle \left( v_{\rm{p,g}} - \langle
v_{\rm{p,g}}\rangle \right)^{2}\rangle}$ (notice that $S_{v}^{2}\neq
S_{x}^{2}+S_{y}^{2}+S_{z}^{2}$). The dotted curve corresponds to the
analytical solution of V\"{o}lk et al. (1980) and Cuzzi et al. (1993)
for $S_{v}$. The solid and dashed curves correspond, respectively, to the radial and
azimuthal solutions of Youdin and Lithwick (2007). With no gas drag, the particles would follow epicyclic paths with the
ratio $\sigma_{v_{x}}^{\rm{p}}/\sigma_{v_{y}}^{\rm{p}}$ of the
radial and azimuthal velocity dispersions equal to 2. Larger ratios
$\sigma_{v_{x}}^{\rm{p}}/\sigma_{v_{y}}^{\rm{p}}$ of around 2.3 measured in the MHD calculation are due
to weak drag forces from the turbulence. Separate numerical
integration of the particle equation of motion~(\ref{eq:partmotion})
with a sinusoidally time-varying gas velocity reproduces the
anisotropy, given variation periods of one to two orbits. Fourier
analysis of the gas velocities at the particle locations in the MHD
calculation indicates the strongest modes do have periods longer than
one orbit. Our neglect of the vertical component of gravity appears
to have little effect on the ratios between particle velocity components, as similar
anisotropy was observed in stratified shearing box calculations
(Carballido et al. 2006), in which for the $\Omega
t_{\rm{s}}$=10 case, $\sigma_{v_{x}}^{\rm{p}} \sim 2.5\sigma_{v_{z}}^{\rm{p}}$.

It is important to remark that the analytical models
of Fig. 2 were originally derived assuming an
isotropic turbulent velocity field, so in principle one
should not expect that they be an accurate description of particle
motion in MRI turbulence. Nevertheless, the particle data obtained
from the simulations conform reasonably well to the models. This will
be discussed below.

\subsection{Pair-wise relative velocities}
\begin{figure*}\label{fig:pair1}
\begin{center}
\includegraphics[width=0.85\textwidth]{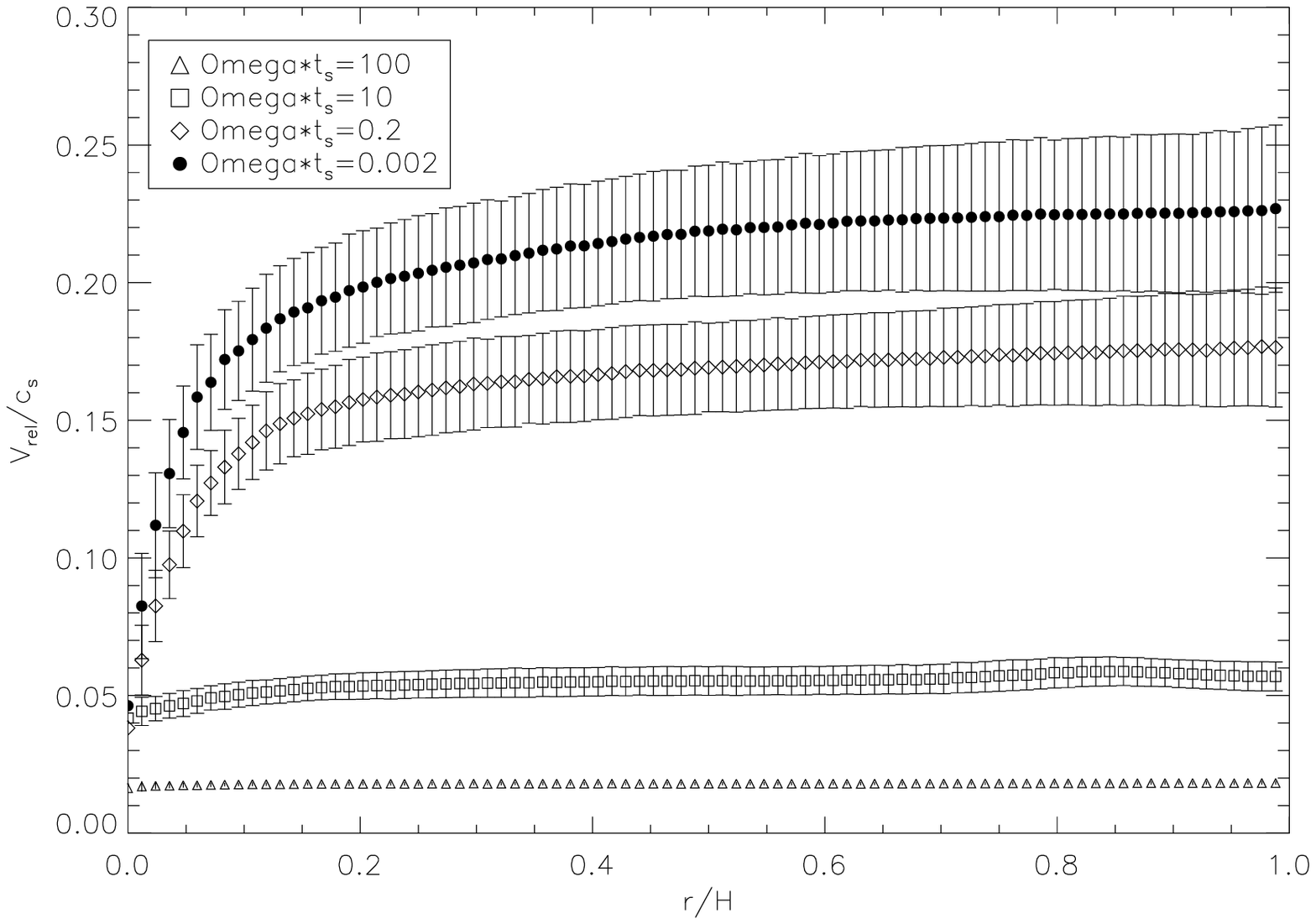}
\end{center}
\caption{Relative speed (in units of the sound speed) as a function of
inter-particle separation (in units of scaleheight),
for pairs of particles with the stopping times shown on the upper left-hand corner. The number
of particles per stopping time is 700. Data have
been binned with respect to $r$, with a bin size equal to the size of
a grid cell in the $x$ direction ($\approx 0.01H$). The values of
$v_{\textrm{rel}}$ decrease with small values of the separation.}
\end{figure*}

\begin{figure*}\label{fig:pair2}
\begin{center}
\includegraphics[width=0.85\textwidth]{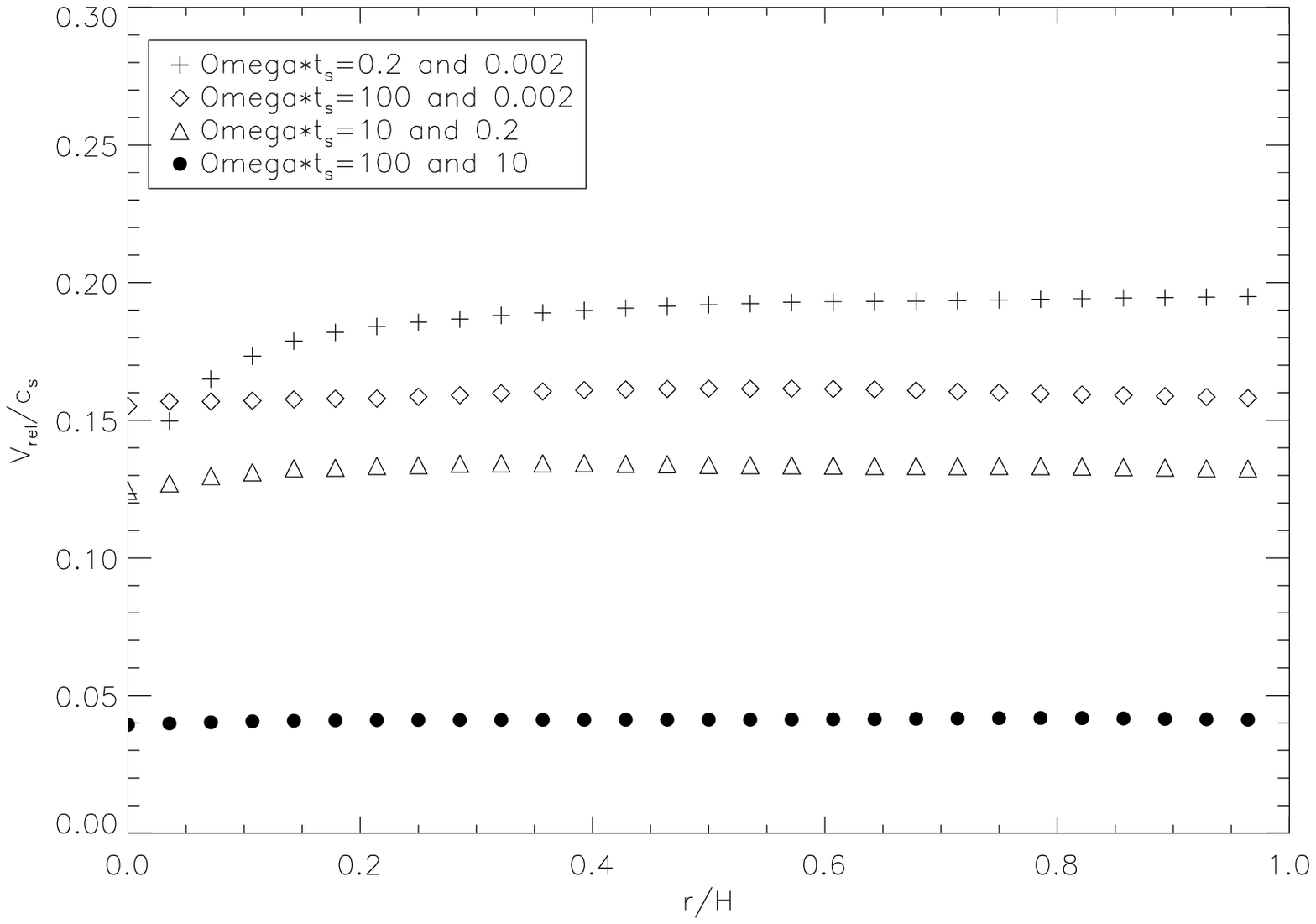}
\end{center}
\caption{Same as Fig. 3, but for pairs of particles with different
stopping times.}
\end{figure*}

Fig. 3 shows the magnitude $v_{\textrm{rel}}$ of the relative
velocity of pairs of particles that are located inside a cubic
sub-volume in the box, in simulation set A, as a
function of inter-particle separation $r$ (in units of the box size
$H$). The velocity has been normalised to the gas sound speed. The sub-volume has dimensions $1\times 1
\times 1$, and extends along the $y$ direction between the values
$y=2.64$ and $y=3.64$. At each snapshot, a sample of 700 particles,
all with the same stopping time, is randomly
selected in the sub-volume, and the relative speed for each of the 244,650 possible pairs
in the sample is calculated. The relative separation between pair
members is binned, with the bin size equal to the $x$-length of a grid
cell. The values of $v_{\textrm{rel}}$ are then averaged in each bin
over a duration of 7 orbits. The data shown correspond to the
stopping times $\Omega
t_{\textrm{s}}$=100 (\textit{triangles}),
10 (\textit{squares}), 0.2 (\textit{diamonds}), and 2$\times 10^{-3}$
(\textit{filled circles}). To calculate
$v_{\textrm{rel}}$, the contribution of the shear has been subtracted
from the $y$ component of the particle velocities. 

It can be seen that at small separations the relative speed decreases appreciably.
This is to be expected, since grains that are
equally coupled to the gas move coherently with small turbulent
eddies. In the limit of particle stopping times less than the lifetime
of the smallest eddies, the laminar flow on scales below the
dissipation scale does not contribute to the random relative grain
velocity. Formally, this can be expressed for grains of stopping times
$t_{\textrm{s}_{1}}$ and $t_{\textrm{s}_{2}}$ as (Weidenschilling 1984)

\begin{equation}\label{eq:vrel}
v_{\textrm{rel}}\sim \frac{u}{\tau}\lvert t_{\textrm{s}_{1}}-
t_{\textrm{s}_{2}} \rvert
\end{equation}   

\noindent where $u$ and $\tau$ are the velocity and lifetime of the
smallest eddy, respectively.

A similar calculation as that of Fig. 3 is presented in Fig. 4,
but for pairs of particles with different stopping times. Each 
symbol represents a different pair of stopping times (the error
bars have been ommited for clarity, but give an uncertainty of $\sim
10\%$ for the data represented by the plus signs, the diamonds and the
triangles, and of $\sim 5\%$ for the filled circles). The relative 
speeds do not decrease as steeply with
decreasing separation as in the case of like particles. Again, from
the order-of-magnitude estimate (\ref{eq:vrel}), $v_{\textrm{rel}}$
should acquire a finite value in the limit of small stopping
times. For the particles considered, this value is a significant
fraction of the sound speed ($\sim 12-15\%$). 

It is worthwhile to point out that as the inter-particle separation
decreases, the effects of numerical resolution may become important
for the purpose of calculating the relative velocities. The size
$\ell$ of the smallest turbulent eddies, to which the smallest
particles are coupled, varies with viscosity $\nu$ as
$\ell \sim \nu^{3/4}$. The viscosity in turn can be affected by the grid
resolution (HGB).

\subsection{Spatial distribution}
In order to
improve the statistics of particle counts, we  have taken as the unit of
volume a group of $3\times 3 \times 3$ grid cells,
referred to as a 3-macrocell. In this way, the average occupation number
is 100 particles per 3-macrocell, with an associated standard
deviation of 10\%.

The placement of the dust particles in the
shearing box, at time $t$=10 orbits, is random, and therefore follows a Poisson
distribution. This may no longer be true at 
subsequent times, when the turbulence has mixed the particles in such
a way that the initial stochastic distribution has been
lost. This situation is apparent in Fig. 5, where the
left panel shows the cumulative distribution function, as a solid
line, of the particle number $n$ per 3-macrocell at the moment when
the particles are introduced in
simulation set B, for the $\Omega t_{\textrm{s}}$=2 case. The dotted line is the
corresponding cumulative Poisson distribution with the same mean
($\langle n\rangle$=100) as the $n$ distribution. The two are alike,
indicating that the initial particle positions are indeed drawn from a
random sample. However, the right panel, which corresponds
to $t$=30 orbits, makes it evident
that the spatial distribution of the particles deviates from a Poisson
distribution. The vertical dashed line in each panel
indicates the point at which the distance $d$ between the two
distributions is greatest. This distance constitutes a measure of the
deviation of one data set from the other. In the particular case shown
in Fig. 5, at $t$=10 orbits the maximum distance is
$d$=0.026, whereas at $t$=30 orbits, $d$=0.55. The situation is
similar at all other times $t>$0, and for the three other hydrodynamic
couplings of set B.

\begin{figure*}\label{fig:kstest}
\begin{center}
\includegraphics[scale=0.5]{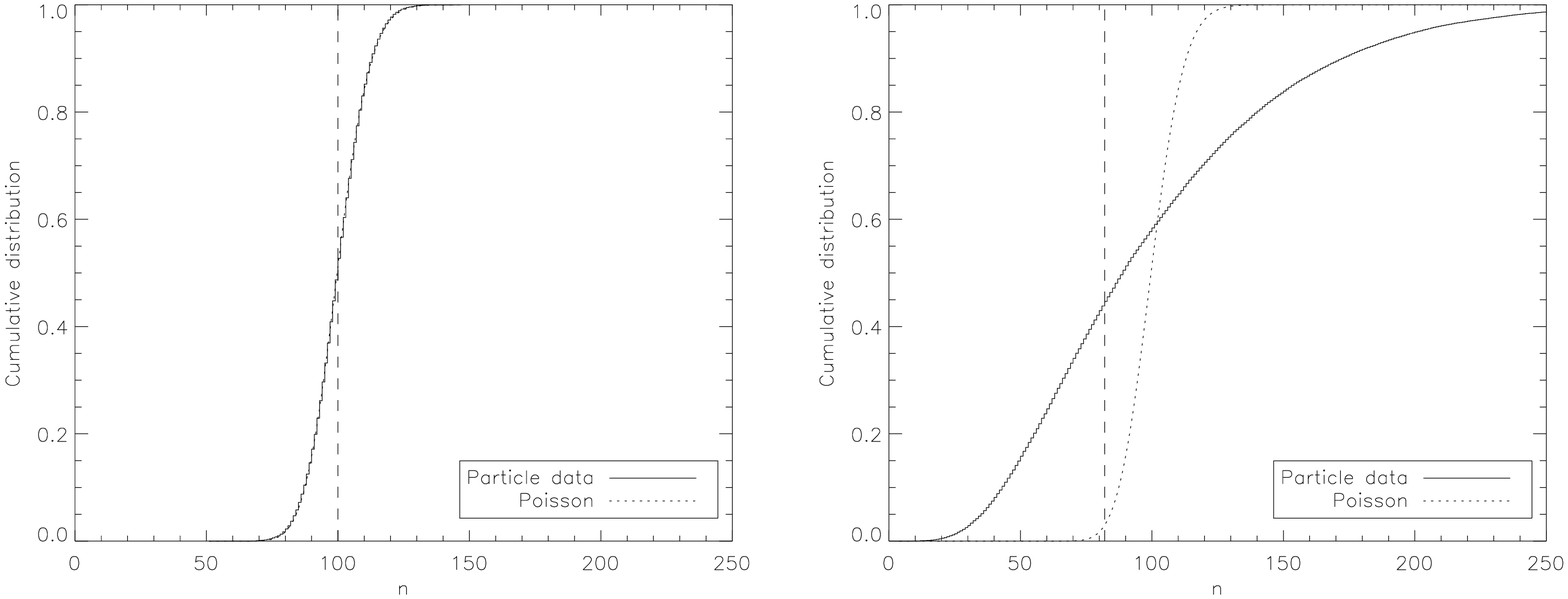}
\caption{Cumulative distributions of the number of particles per
3-macrocell, for $\Omega t_{\rm{s}}=2$. The left panel represents the
initial distribution at $t$=10 orbits. The solid line corresponds to
the particle distribution $n$, while the dotted line is the corresponding
Poisson distribution with the same mean. The vertical dashed line indicates
the value of $n$ at which the vertical separation $d$ between the two
distributions is greatest. The right panel shows that the particle
distribution is not Poissonian at $t$=30 orbits. This is true in
general for $t>$0 and the three other values of $\Omega t_{\rm{s}}$.}
\end{center}
\end{figure*}

To measure the amount of solid particle clumping in the shearing box,
we calculate the number of particles in each macrocell and divide by its
mean:

\begin{equation}\label{eq:clumping}
C\equiv \frac{n}{\langle n \rangle}
\end{equation}

\begin{figure*}\label{fig:snapclumpiness}
\begin{center}
\includegraphics[width=0.85\textwidth]{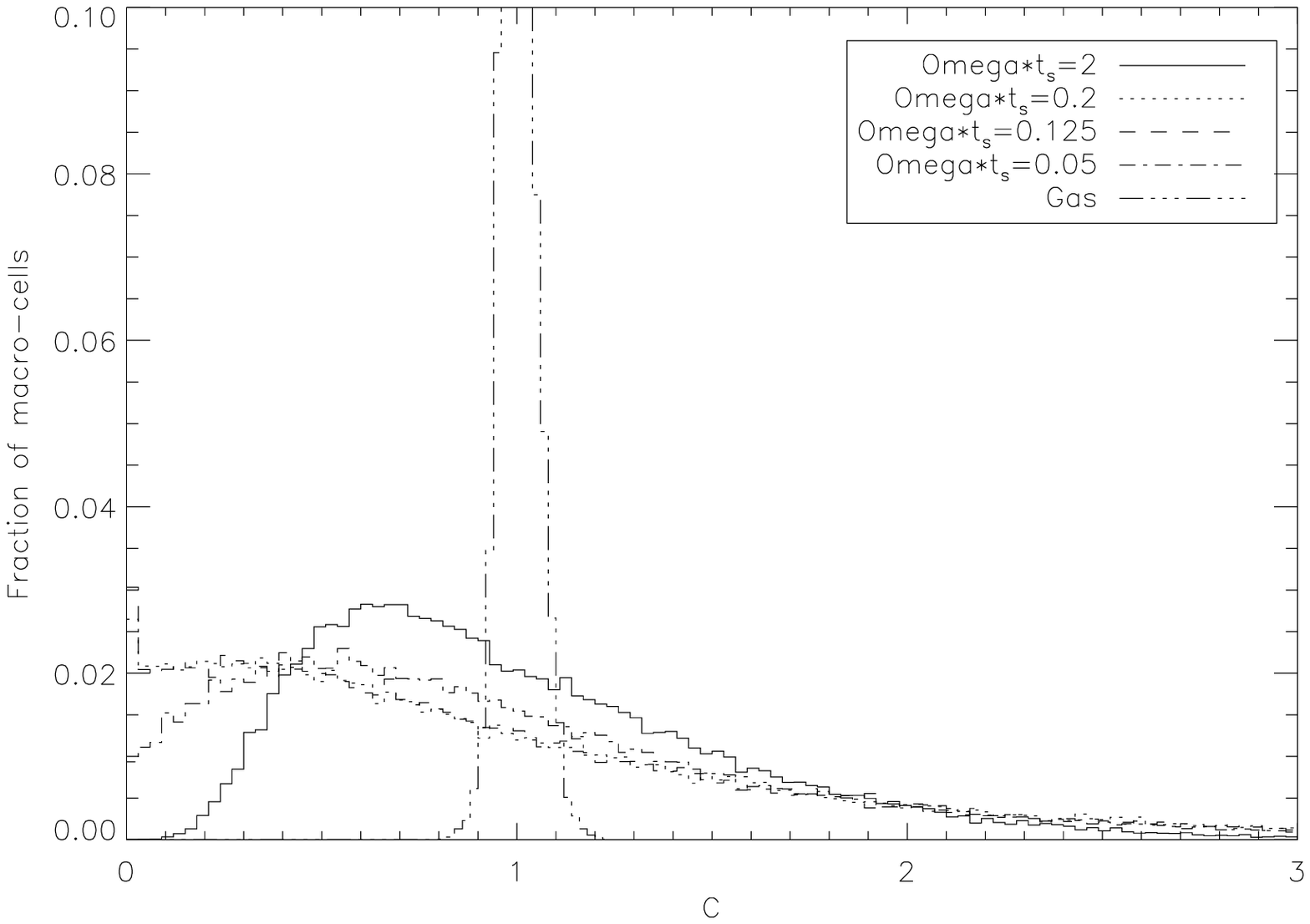}
\caption{Distribution of the clumping number $C\equiv n/\langle n
\rangle$ at $t=$20 orbits, for all particle stopping times. All solids exhibit an
overabundance of 3-macrocells for which $n$ is larger than the
mean. All stopping times are more clumped than the gas (\textit{dash-tripple dotted line}).}
\end{center}
\end{figure*}

 Fig. 6 shows distributions of $C$
for each particle-gas coupling, at $t$=20 orbits. These distributions
are typical throughout the duration of the simulation. Also shown for comparison is the
distribution of gas clumping, $\rho_{\rm{g}}/\langle
\rho_{\rm{g}}\rangle$. Notice that the
solids exhibit more over-densities with respect to
the mean ($\langle C \rangle=1$) than under-densities, showing a considerably high number of
macrocells that have values of $C$ above the average occupation number
$\langle n \rangle$.
The highest $C$ attained is $C_{\rm{max}}$=26.5, by the particles with $\Omega
t_{\rm{s}}$=0.125, at $t=$27.1 orbits. A history of $C_{\rm{max}}$ is
plotted in Fig. 7, where the time axis has been shifted to the instant
at which the particles are introduced.

\begin{figure*}\label{fig:maxclumphist}
\begin{center}
\includegraphics[width=0.85\textwidth]{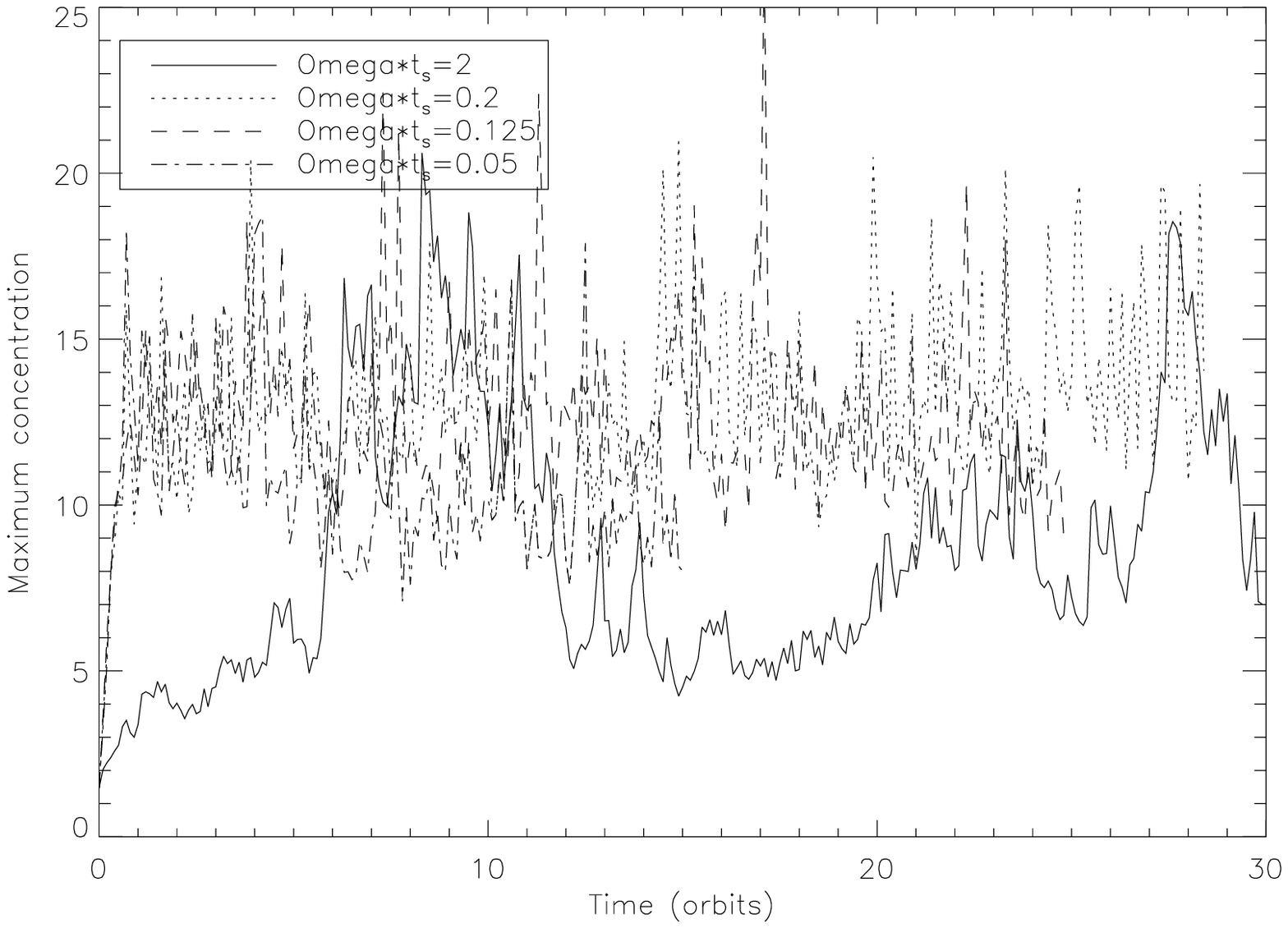}
\caption{Time history of the maximum clumping factor $C_{\rm{max}}$
for the particle stopping times used in run B. The time marked as
$t=0$ orbits corresponds to the instant at which particles are
introduced in the flow.} 
\end{center}
\end{figure*}

\begin{figure*}\label{fig:avclumpiness}
\begin{center}
\includegraphics[width=0.85\textwidth]{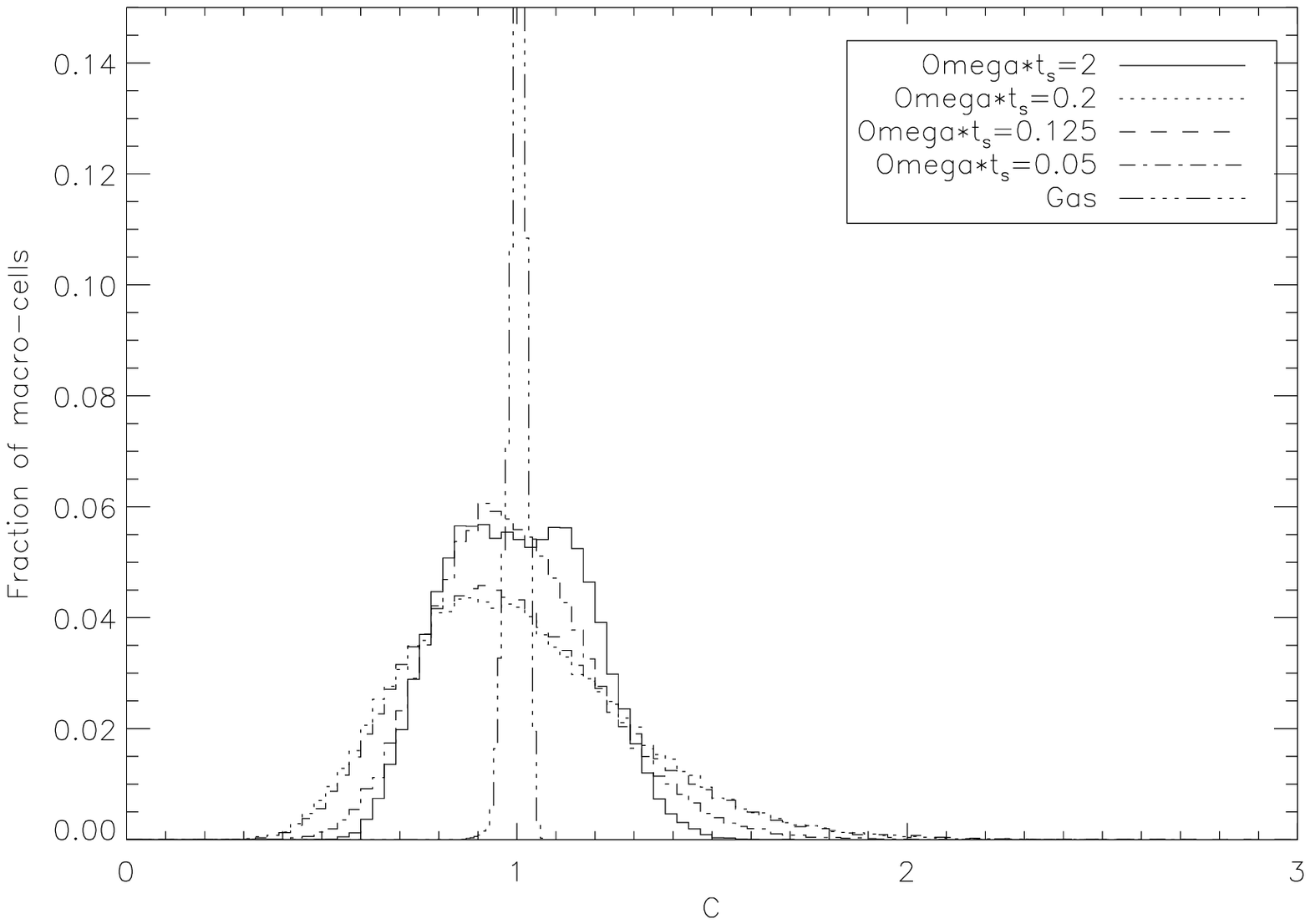}
\caption{Time-averaged distribution of the clumping number $C$, for
all particle stopping times. The tails of the distributions, although
not particularly long, are indicative of the non-Poissonian nature of
the number density counts.}
\end{center}
\end{figure*}

The time-averaged $C$ distributions are shown in
Fig. 8. Note the difference in the shapes of the
distributions compared to those in Fig. 6,
particularly for $\Omega t_{\rm{s}}=0.2$, 0.125 and 0.05 at $C=0$. This can be
understood by considering that the time-averaging procedure
effectively assigns a non-zero number of particles to each
3-macrocell, which at a specific snapshot (but not all) is likely to
be empty. The non-Poissonian
character of the particle spatial distribution is also evident in
Fig. 8, as the distributions
possess appreciable ``tails'' extending in the direction of increasing $C$. This
is normally associated with spatial correlations in a system of
particles in which their positions are not statistically independent
(Landau \& Lifshitz 1980). In the case of particles
immersed in turbulence, these correlations arise from the
intermittency of turbulent mixing. The correlations are manifested
by an increase in the variance of the particle counts, with respect to
a purely random distribution.  

Positive correlations can
induce clustering at some range of spatial scales. One way to quantify
this clustering is by means of the pair correlation function
 
\begin{equation}\label{eq:paircorrel}
\eta(\Delta r)=\frac{\langle n(r)n(r+\Delta
r)\rangle}{\langle n(r) \rangle^{2}}-1
\end{equation}

\noindent where the particle number density $n$ is evaluated at two 
points separated by the distance $\Delta r$. The measurement of this
quantity is common in the study of cloud droplet clustering in the
presence of turbulence in the
Earth's atmosphere (Kostinski \& Jameson 2000). In order to
calculate $\eta$ for the population of solids in our shearing box, we
proceed as follows. Given the anisotropic character of the turbulent
eddies present in the box (HGB), $\eta$ is measured independently
along each spatial direction $i$ (=1,2,3, or $x,y,z$):

\begin{equation}\label{eq:paircorrelx}
\eta_{i}(\Delta x_{i})=\frac{\langle n(x_{i})n(x_{i}+\Delta
x_{i})\rangle}{\langle n(x_{i}) \rangle^{2}}-1
\end{equation}

Furthermore, to obtain a sufficient number of points along each direction, at which to
evaluate $n$, we consider 2-macrocells instead of the 3-macrocells
used to calculate the clumping factor $C$. To avoid the effect of the
periodic boundary conditions in the $y$ and $z$ directions, as well as
the quasi-periodic conditions along the radial direction $x$, the separation $\Delta x_{i}$ between 2-macrocells is
constrained to be, at most, half of the corresponding box length. Thus, for example, to
evaluate $\eta_{x}$ at separations of one macrocell, a horizontal row of grid cells is chosen, say
one with end cells at $x=-0.5$ and $x=0.5$, and with fixed $y$ and $z$
coordinates. The particle number density $n$ is
measured at the first 2-macrocell in this row, and also at the next
macrocell. The product of these two values of $n$ is stored, and the
process is repeated starting at the second macrocell. If $m$
is the number of 2-macrocells in the row, the process ends with the
product of $n$ evaluated at macrocells $m/2-1$ and $m/2$. The average
of all the products is calculated and inserted in the right-hand side of
Eq. (\ref{eq:paircorrelx}). 

The row of grid cells used in the calculation just described is
located at specific $y$ and $z$, and it is only one of 84 $\times 180=15,120$ similar
horizontal rows. With this number of values for $\eta_{x}(\Delta x)$,
an average over all rows at all $y$ and $z$ gives the volume-averaged
pair correlation function $\langle \eta_{x}(\Delta x)\rangle$. The
functions $\langle \eta_{y}(\Delta y)\rangle$ and $\langle
\eta_{z}(\Delta z)\rangle$ are obtained in an analogous manner. 

The volume-averaged pair
correlation functions for the $\Omega t_{\rm{s}}$=0.05 particles, at
the time of their introduction in the flow ($t=10.0$ orbits), are identically zero,
reflecting the statistical independence of the initial particle numbers. This
also holds for the cases $\Omega t_{\rm{s}}$=0.125, 0.2 and 2. The
measured $\langle \eta_{i}\rangle$ at the end of the simulation are shown in
Fig. 9. The columns correspond to the $x$,
$y$ and $z$ directions, and the rows represent, from top to bottom,
the stopping times $\Omega t_{\rm{s}}=$2, 0.2, 0.125 and 0.05. The
error bars correspond to the data dispersion at each separation
$\Delta x_{i}$. 

\begin{figure*}\label{fig:paircorrel_2orbs}
\begin{center}
\includegraphics[scale=0.5]{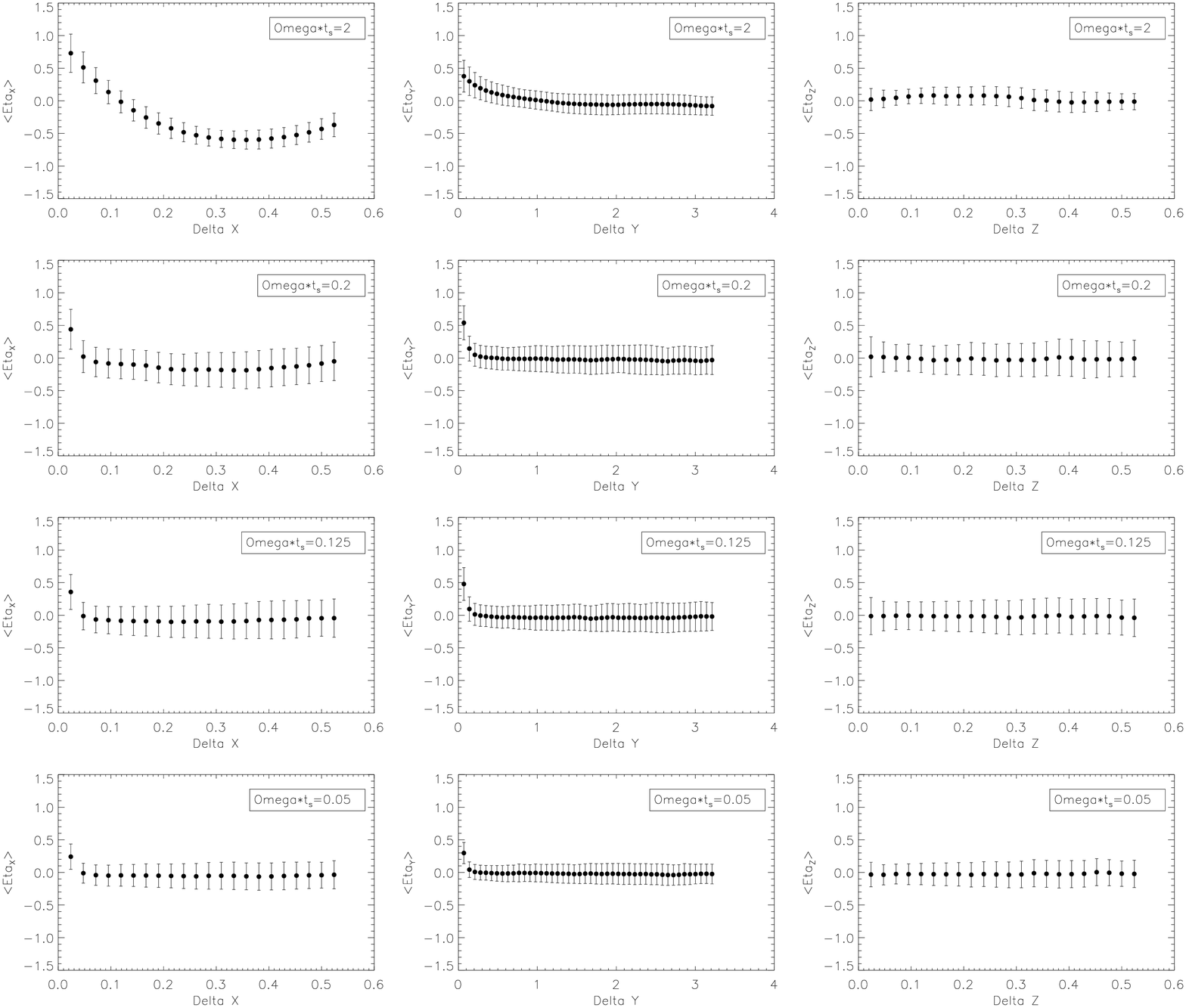}
\caption{Pair correlation function for all particle
stopping times, at the end of simulation set B. The left, centre and right columns
correspond to the $x$, $y$ and $z$ directions, respectively. First
row:  $\Omega t_{\rm{s}}=2$; second row: $\Omega t_{\rm{s}}=0.2$; third
row: $\Omega t_{\rm{s}}=0.125$; fourth row: $\Omega
t_{\rm{s}}=0.05$. The vertical correlation functions are very nearly
zero, indicating that particles are evenly mixed in that direction.}
\end{center}
\end{figure*}

One feature is immediately noticeable from
Fig. 9: the vertical correlation
function (third column) is very nearly flat for the four values of
 $\Omega t_{\rm{s}}$. This would seem to indicate that particles
remain almost evenly mixed in the vertical direction, without deviating much from
the initial random distribution. The particles show
concentrations over radial distances of approximately 0.1$H$ (for $\Omega
t_{\rm{s}}$=2) and 0.05$H$ (for the other stopping times). In the
azimuthal direction the respective correlation lengths are $\approx H$
and $\approx 0.3H$. This range of distances represents between 5\% and 16\% of the box size. 

The largest correlation lengths belong to the $\Omega t_{\rm{s}}$=2
particles, which also show the lowest average concentration (Figs. 7
and 8). Even taking the highest value of $C_{\rm{max}}$ for this
stopping time ($\approx$21), the surface density reached in a clump of
dimensions $\langle \eta_{x}\rangle \times \langle \eta_{y}\rangle$ is
$\approx 2.1\times 10^{4} H^{-2}$, compared to $2.65\times 10^{5}
H^{-2}$ for the most highly concentrated particles, those with $\Omega
t_{\rm{s}}$=0.125.

In order to constrain the lifetime of typical particle associations from the
simulation data, it is necessary to refine the working definition
of a clump. Here a clump is understood to be a group of particles that
has a size smaller than the length $\delta$ of the diagonal of a
3-macrocell. The size of the group is determined by calculating a
clump ``diameter'', based on the maximum and minimum values of the
particle coordinates. 

For each value $\Omega t_{\rm{s}}$=0.2, 0.125 and 0.05, the clump with
the highest
particle density was followed during subsequent snapshots. In all
cases, the particles become detached from the original
clump after $\Delta t= 0.1$ orbits, in such a way that a calculation
of the new diameter yields a value larger than $\delta$. The fact that
$\Delta t$ is
the interval between
snapshots indicates that the lifetimes are less
than what is possible to probe using the available data. The situation
is similar if by clumps it is meant groups with horizontal dimensions
corresponding to the particle correlation length.

\section{DISCUSSION}
The measurement of solid particle velocity dispersions in MRI turbulence allows
to test the analytic model of V\"{o}lk et al. (1980) and Cuzzi et
al. (1993), which adopted a Kolmogorov energy spectrum. The turbulence
that develops in the shearing box setup that we use follows roughly a
Kolmogorov scaling (HGB), and the dispersion of the particle velocity
\textit{magnitude} is generally consistent with the previous results. 

The radial and azimuthal velocity dispersions were found to conform to
the analytical results of Youdin and Lithwick (2007), who use a
turbulence model in which gas velocity correlations $\langle
v_{x}v_{y}\rangle$ are weak compared to $\langle
v_{x}^{2}+v_{y}^{2}\rangle$. This is indeed the case in the
MRI-generated turbulent flow present in our shearing box.

Recent analytical calculations have provided closed-form expressions
for the relative velocity between solids of different sizes, as a
function of stopping time, in isotropic turbulence (Ormel \&
Cuzzi 2007). Those results validate the use of Eq.~(\ref{eq:vrel}) in
a regime where the stopping times of both particles are smaller than
the lifetime $\tau$ of the smallest eddies. From a measurement of the
correlation time of the gas velocities, we find $\tau\sim 0.1$ orbits.
Therefore, Eq.~(\ref{eq:vrel}) will not apply to all the stopping
times shown in Fig. 4, since the stopping times for the $\Omega
t_{\rm{s}}$=10 and $\Omega t_{\rm{s}}$=100 particles are, respectively,
$\sim$1.6 and $\sim$16 orbits. Nevertheless, a separate calculation
shows that relative velocities between pairs of the three smallest stopping
times ($\Omega t_{\rm{s}}$=0.002, 0.02 and 0.2, which correspond to
$\sim 3 \times 10^{-4}, 3 \times 10^{-3}$ and $3 \times 10^{-2}$
orbits) are well described by Eq.~(\ref{eq:vrel}) at short
separations, with the lowest velocity ($v_{\rm{rel}}\approx
0.09c_{s}$) corresponding to pairs with $\Omega t_{\rm{s}}$=0.002 and
0.02. 

For $t_{\rm{s}}> t_{L}$, with $t_{L}$ being the overturn time of
the largest eddies (typically taken as one orbital period), the model
of Ormel \& Cuzzi (2007) gives

\begin{equation}\label{eq:ormelcuzzi}
v_{\rm{rel}}^{2}=v_{\rm{gas}}^{2}\left(\frac{1}{1+\Omega t_{\rm{s_{1}}}}+\frac{1}{1+\Omega t_{\rm{s_{2}}}}\right)
\end{equation}

\noindent where $v_{\rm{gas}}$ is the velocity of the
turbulent flow, and it is assumed that $t_{\rm{s_{1}}}>t_{\rm{s_{2}}}$
whitout loss of generality. Direct measurement from our simulation
data yields $v_{\rm{gas}}\approx 0.12c_{\rm{s}}$, and so for the two
largest stopping times studied ($\Omega t_{\rm{s}}$=10 and $\Omega
t_{\rm{s}}$=100) we obtain $v_{\rm{rel}} \approx 0.04c_{\rm{s}}$, in close
agreement with the numerical results of Fig. 4 (filled circles).

The two-point correlation function calculated in Section 3.4 gives an
estimate of the length scale over which particle clustering occurs. In
a MMSN scenario at 5 AU, in which the scale height $H$ of the disc is
$\sim 4\times 10^{12}$ cm, the minimum correlation length of the particle number
density would be $\xi \sim 0.05H\sim 2 \times 10^{6}$ km. A recent
model of chondrule formation, in which
molten chondrules come to equilibrium with the gas evaporated from
other chondrules, indicates that these particles must have originated
in regions larger than 6,000 km in radius, or $\xi \sim 6\times 10^{-4}H$ (Cuzzi \& Alexander
2006). The smallest particles in our study of the spatial
distribution correspond roughly to solids with a radius of 20 cm, two
orders of magnitude larger than typical chondrule sizes. It would be
necessary to perform additional calculations with particle stopping
times that correspond to this size range, in order to further
constrain the value of $\xi$. However, it is unlikely that current computational
resources would allow for the necessary resolution to probe such small
correlation lengths.

We have estimated the
Jeans radius $R_{J}$ of the maximally concentrated clump for each stopping
time, following calculations by JKH. Using similar
disc parameters (column density $\Sigma=900$ g cm$^{-2}$, average
dust-to-gas density ratio $\zeta=0.01$, and scale height-to-radius
ratio $H/r=0.04$) and a turbulent diffusion coefficient
$\delta_{\rm{t}}=\alpha=8\times 10^{-3}$ inferred from
our shearing box model, the values of
$R_{J}$ are smaller than the typical
clump size by factors between $\sim$9 and $\sim$250. The clump radius
that is closest to $R_{J}$ corresponds to the $\Omega t_{\rm{s}}$=2
particles. Under these assumptions, our dense particle associations would
not be subject to a gravitational collapse. 

Even though the stopping times used in this calculation and those
studied by JKH are different, we can still compare our results for $\Omega
t_{\rm{s}}$=2 with their $\Omega t_{\rm{s}}$=1 case. The main
difference resides in the particle velocity dispersion, which we
measure as $0.036c_{\rm{s}}$, approximately $50\%$ larger than they
obtain. If our maximally concentrated $\Omega t_{\rm{s}}$=2 group were
to be gravitationally unstable, the velocity dispersion would need to
be a factor $\sim$ 2.5 smaller. We also note that the radial drift
incorporated by JKH leads to particle concentrations that are stronger
than in the cases presented here (Johansen et al. 2007), making
gravitational collapse more likely.

\section{CONCLUSIONS}
We have performed numerical MHD calculations of solid particle velocities
and spatial distribution in a turbulent protoplanetary disc, in the
context of the shearing box model. The results suggest that, even
though the turbulent flow has an anisotropic spectrum, the dispersion
of the particle velocity magnitude (normalised by the corresponding
dispersion for the gas) as a function of particle stopping
time is consistent with analytical models that assume
isotropy. Nevertheless, the dispersions of the individual velocity
components for each stopping time are different. In the case of
particles with $\Omega t_{\rm{s}} \gtrsim 1$, the radial
component of the solid-to-gas velocity dispersion ratio is larger than the azimuthal and vertical
components typically by factors of $\sim 2.3 - \sim$ 2.5. The relative speed
between pairs of particles with the same stopping time decreases
towards small values at small separations,
whereas for pairs of different stopping times the relative speed
reaches a finite value that depends on the sizes of the pair members.

The clustering induced on the particle spatial distribution by the
turbulent flow gives rise to short-lived clumps. Under
conditions of a minimum-mass solar nebula, these clumps are at least
nine times larger than their respective Jeans radii, as a result of
high particle velocity dispersions. Gravitational
collapse would therefore seem to be an unlikely mechanism in the
formation of more massive objects, under the conditions that we assumed.

Future calculations should increase the numerical resolution used in
the measurement of the relative velocities between particles, in order
to improve fittings to existing analytical models.

\bigskip

\section*{ACKNOWLEDGMENTS}
This work benefited from discussions with Jim Pringle, Richard
Nelson, and Gordon Ogilvie. Suggestions from an
anonymous referee were most valuable. AC acknowledges support
from CONACYT scholarship 167912 and a Caltech Postdoctoral Scholarship. Part of
this work was carried out at the Jet Propulsion Laboratory, which is
operated by the California Institute of Technology under contract to NASA.

\end{document}